\documentclass[twocolumn,a4paper,10pt,noshowpacs,prl,aps,superscriptaddress,amsmath,amssymb,amsfonts,floatfix]{revtex4}
\pdfoutput=1
\usepackage[utf8x]{inputenc}
\usepackage{graphicx,array}
\usepackage{rotating}
\usepackage[caption=false]{subfig}
\usepackage[normalem]{ulem}
\usepackage[pdftex]{hyperref}
\hypersetup{colorlinks=true,citecolor=blue,linkcolor=red,urlcolor=blue}
\usepackage[all]{hypcap}
\usepackage{color}

\begin{document}

\title{Graphene on {\it h}-BN: to align or not to align?}

\author{R.\ Guerra}\email{guerra@sissa.it}
\affiliation{International School for Advanced Studies (SISSA), Via Bonomea 265, 34136 Trieste, Italy.}
\affiliation{Dipartimento di Fisica, Universit\`a degli Studi di Milano, Via Celoria 16, 20133 Milano, Italy}

\author{M.\ M.\ van Wijk}
\affiliation{Institute for Molecules and Materials, Radboud University Nijmegen, Heyendaalseweg 135, 6525 AJ Nijmegen, The Netherlands.}

\author{A.\ Vanossi}
\affiliation{CNR-IOM Democritos National Simulation Center, Via Bonomea 265, 34136 Trieste, Italy.}
\affiliation{International School for Advanced Studies (SISSA), Via Bonomea 265, 34136 Trieste, Italy.}

\author{A.\ Fasolino}
\affiliation{Institute for Molecules and Materials, Radboud University Nijmegen, Heyendaalseweg 135, 6525 AJ Nijmegen, The Netherlands.}

\author{E.\ Tosatti}
\affiliation{International School for Advanced Studies (SISSA), Via Bonomea 265, 34136 Trieste, Italy.}
\affiliation{The Abdus Salam International Centre for Theoretical Physics (ICTP), Strada Costiera 11, 34151 Trieste, Italy.}

\begin{abstract}
The contact strength, adhesion and friction, between graphene and an incommensurate crystalline substrate such as {\it h}-BN depends on their relative
alignment angle $\theta$. The well established Novaco-McTague (NM) theory predicts for a monolayer graphene on a hard bulk {\it h}-BN crystal face
a small spontaneous misalignment, here $\theta_{NM}$\,$\simeq$\,0.45 degrees which if realized would be relevant to a host of electronic properties
besides the mechanical ones.
Because experimental equilibrium is hard to achieve, we inquire theoretically about alignment or misalignment by simulations based on dependable
state-of-the-art interatomic force fields. Surprisingly at first, we find compelling evidence for $\theta = 0$, i.e., full energy-driven alignment
in the equilibrium state of graphene on {\it h}-BN.
Two factors drive this deviation from NM theory. First, graphene is not flat, developing on {\it h}-BN  a long-wavelength out-of-plane corrugation.
Second, {\it h}-BN is not hard, releasing its contact stress by planar contractions/expansions that accompany the interface moir\'e structure.
Repeated simulations by artificially forcing graphene to keep flat, and {\it h}-BN to keep rigid, indeed yield an equilibrium misalignment similar to
$\theta_{NM}$ as expected.
Subsequent sliding simulations show that friction of graphene on {\it h}-BN, small and essentially independent of misalignments in the
artificial frozen state, strongly increases in the more realistic corrugated, strain-modulated, aligned state.
\end{abstract}

\maketitle

\section{Introduction}

Graphene, {\it h}-BN, MoS2, and other materials provide strong 2D monolayers of great importance in physics and technology. Practical use of such monolayers
generally requires deposition on a substrate, often a crystal surface. Understanding the alignment, adhesion and friction between the two is instrumental
to that end. The monolayer 2D graphene lattice and that of a substrate such as  {\it h}-BN are generally incommensurate -- not related through
a rational fraction. That situation may lead to structural lubricity (sometimes called superlubricity), involving the possible vanishing of static friction
and smooth sliding in the absence of defects.\cite{vanossi2013} On the other hand, the theory of incommensurate epitaxy, developed long ago in the context of adsorbed
rare gas monolayers by Novaco \& McTague\cite{NM1977,NMPRB1979} and others,\cite{shiba1979, shiba1980} predicted a striking structural
effect -- immediately confirmed experimentally\cite{shaw78} -- consisting of a small spontaneous misalignment angle $\theta$\,=\,$\theta_{NM}$ of
the adsorbed monolayer as a whole relative to the substrate axes. Such misalignment influences the contact strength between lattices with
several consequences including a change of friction, as recently found in a different context.\cite{mandelli2015}

As of now however, a sharp assessment
of the equilibrium alignment or misalignment and of the friction of graphene on pertinent substrates, such as {\it h}-BN, Cu, and others, is missing.
Existing experimental work with {\it h}-BN deposited graphene\cite{zhang2015, woods2016macroscopic, yankowitz2012} is abundant, and a variety
of observed deposition angles are reported.  Generally, it appears that the deposition history and kinetics dominates the relative angle much more
than subtle energy differences connected with misalignment. The precision of the observed self-rotation of micron sized graphene flakes on h-BN
towards small angles after annealing\cite{woods2016macroscopic} is limited to $\theta$\,$<$\,0.7$^\circ$. Most experiments concerning equilibrium
alignment near $\theta$\,=\,0 remain inconclusive in this respect. The question whether the equilibrium graphene geometry is aligned or misaligned
with {\it h}-BN or other substrates must therefore be resolved theoretically.

\begin{figure*}[t!]
 \centering
 \includegraphics[width=0.49\columnwidth]{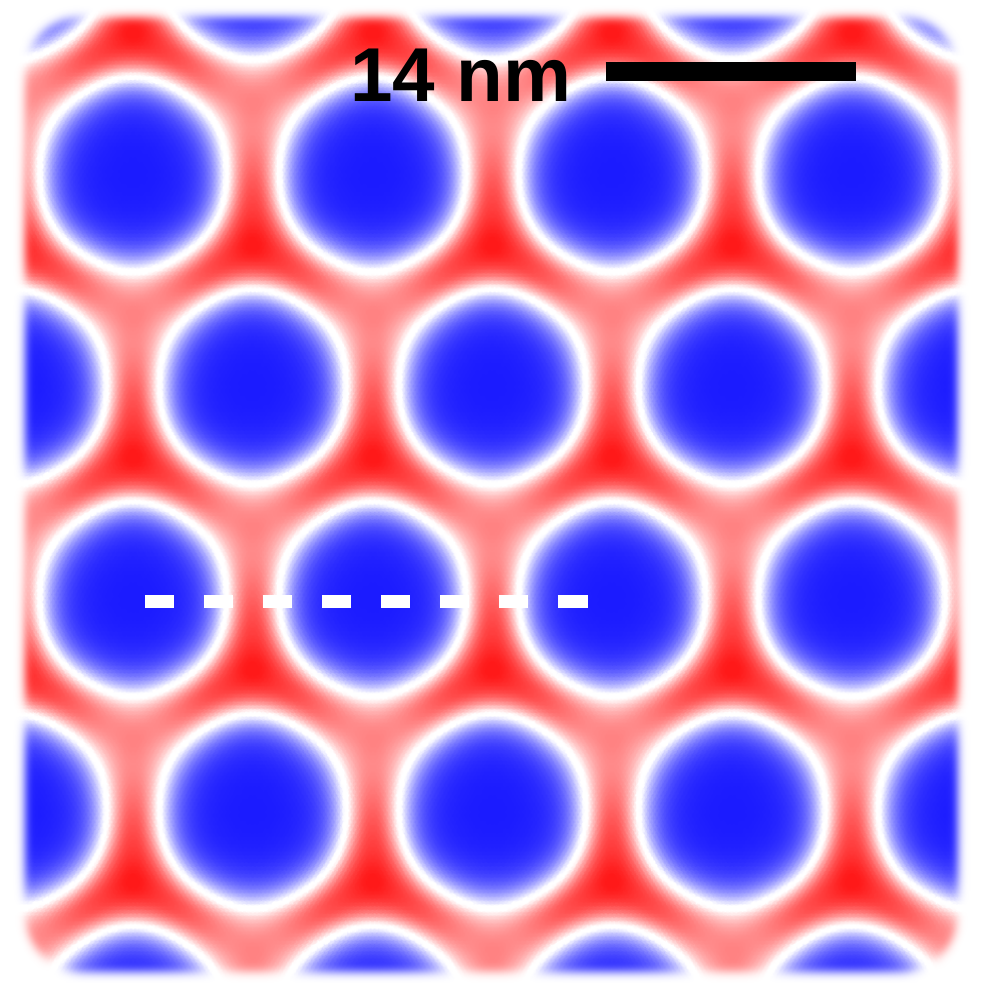}
 \includegraphics[width=0.49\columnwidth]{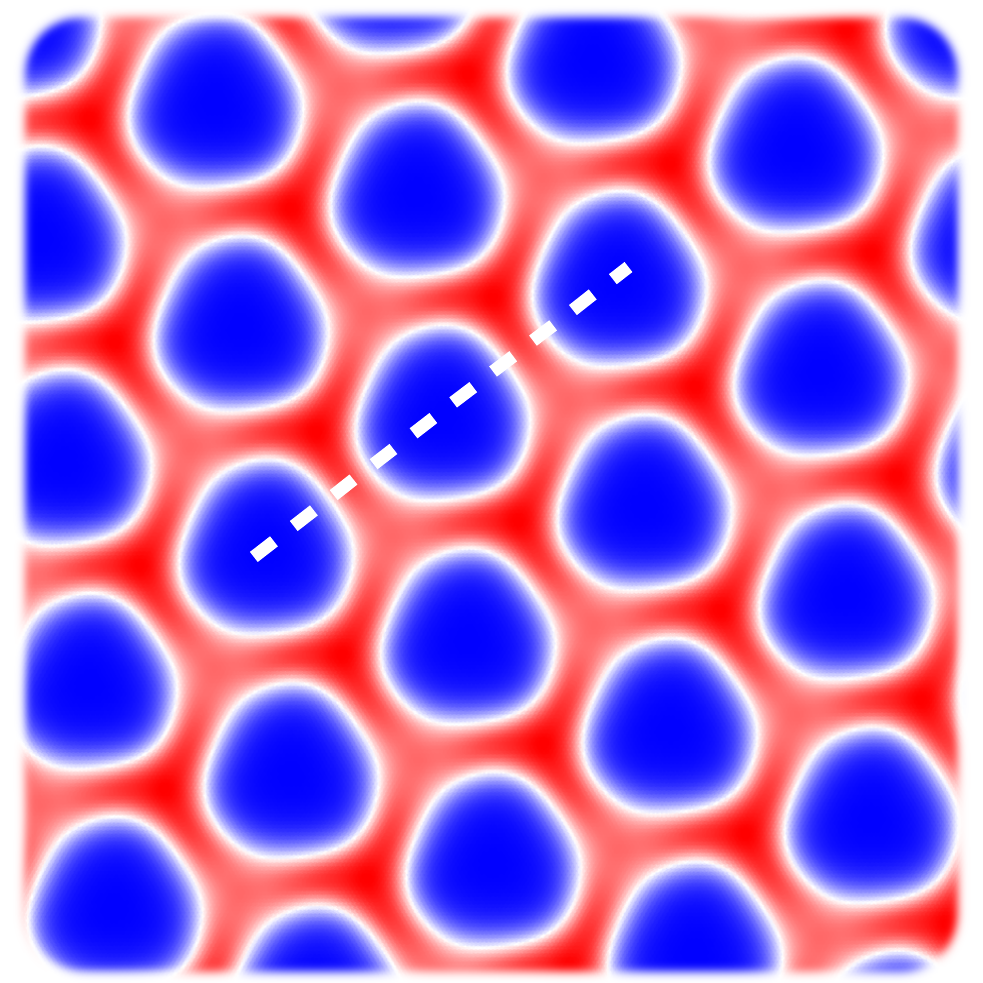}
 \includegraphics[width=0.49\columnwidth]{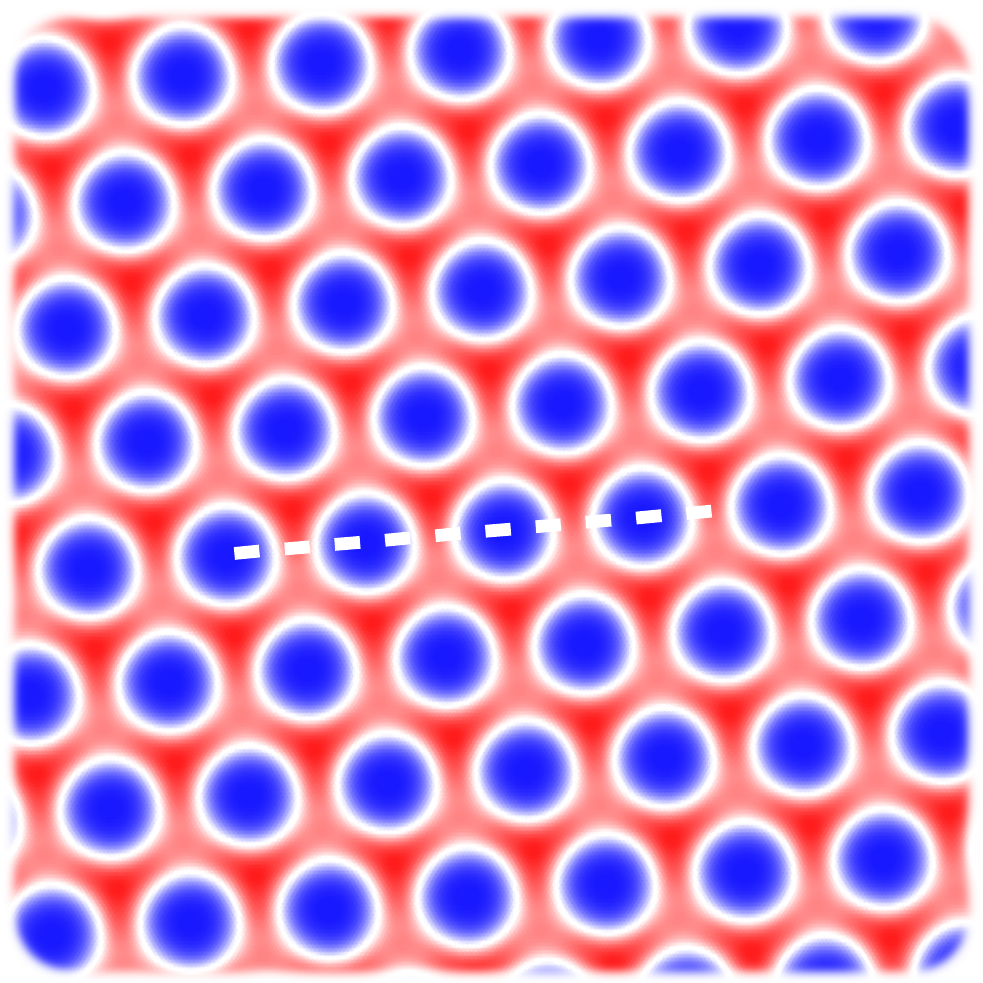}\\
 \includegraphics[width=1.51\columnwidth]{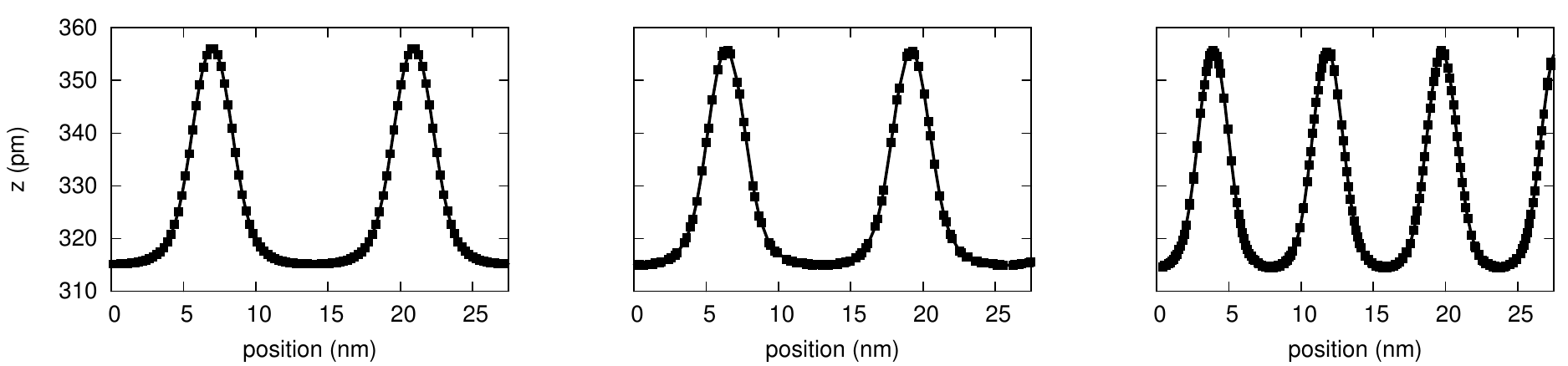}\hspace{6mm}
 \caption{Moir\'e pattern of graphene relaxed on {\it h}-BN at $\theta $\,=\,0 (left panel), 0.45$^\circ$ (center panel), and 1.5$^\circ$ (right panel).
   Blue (red) color corresponds to graphene atoms at inter-layer distance $z$\,=\,315\,pm (363\,pm) from {\it h}-BN. The total system size of each panel
   corresponds to 56$\times$56\,nm$^2$. Calculated $\psi$,$L$ pairs are 0.0$^\circ$,14.0\,nm (left), 23.0$^\circ$,12.8\,nm (center), and 54.7$^\circ$,7.9\,nm (right).
   The z-coordinates of graphene along the dashed traces above the BN plane are reported below each panel; the amplitude of this out-of-plane corrugation agrees,
   quantitatively, very well with experimental findings.\cite{zhang2015,LeRoy2011}}
 \label{fig.moire}
\end{figure*}

According to Novaco-McTague\cite{NMPRB1979} the predicted equilibrium misalignment angle between an adsorbed
monolayer and an underlying substrate lattice is generally nonzero and equal to

\begin{equation}\label{eq.novaco}
 \theta_{NM} = \arccos\left(\frac{1+\rho^2\left(1+2\delta\right)}{\rho\left[2+\delta\left(1+\rho^2\right)\right]}\right)
\end{equation}

where $\rho = a_s / a_C$, with $a_s$ the lattice constant of the
hard
substrate and $a_C$ the lattice constant of the adsorbed layer.
The misalignment depends on the ratio of the sound velocities $c_L$ and $c_T$ of longitudinal acoustic (LA) and
transverse acoustic (TA) phonon modes of the adsorbed layer through the parameter

\begin{equation}\label{eq.soundvel}
 \delta = \left(c_L/c_T\right)^2-1~.
\end{equation}

Importantly, there will only be a misalignment if

\begin{equation}\label{eq.threshold}
\rho \delta > 1~~.
\end{equation}

The assumptions of this theory include a) incommensurate contact; b) weakness of the interaction between the two lattices; c) rigidity of the substrate;
d) flatness of the adsorbate (i.e.,negligible surface-normal displacements of the monolayer), making the problem strictly two-dimensional.
The physics of this misalignment has been clear for a long time. At perfect alignment, $\theta$\,=\,0, the misfit dislocations ("solitons") composing the moir\'e
pattern formed at the adsorbate/substrate contact concentrate into stripes the necessary 2D compression-expansion waves, strictly longitudinal in character and
therefore energetically costly.
Even a small misalignment angle allows the energy balance to change drastically. The moir\'e pattern size shrinks and therefore the soliton density increases: but
the soliton's nature turns at the same time from longitudinal to shear. The latter is energetically cheaper because the shear sound velocity is generally much lower
than the longitudinal one. As soon as parameters in Eq.~\ref{eq.novaco} are such that the elastic energy drop overcompensates the cost, the misalignment becomes
energetically favorable, and is realized in full equilibrium. Soon after its prediction  it was  indeed observed experimentally for Ar monolayers on graphite.\cite{shaw78}
So universal  is this misalignment mechanism that it even occurs for a colloid monolayer in an optical lattice, where characteristic distances are three-four orders of
magnitude larger than for the rare gas adsorbed monolayers for which it was developed.\cite{mandelli2015}

The question we therefore address is: should monolayer graphene, or {\it h}-BN, or MoS2, etc., also exhibit a misalignments by some small
Novaco-McTague-type angle, once
deposited on an incommensurate substrate? The effects of misalignment, if present, should be relevant to friction, which is generally influenced by the mutual
lattice orientation. It would also influence a variety of important physical and technological phenomena from growth to mechanical, electrical and electronic.
Last but not least, misalignement changes the length of moir\'e patterns which is used in experiments to establish the effective lattice mismatch.

\begin{figure}[t!]
 \centering
 \includegraphics[width=0.7\columnwidth]{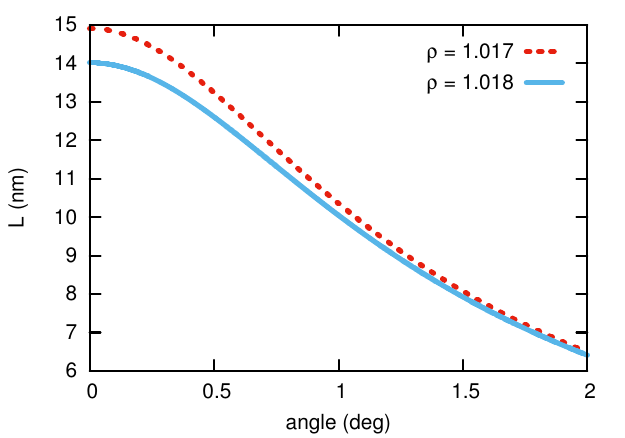}
 \caption{Length of the moir\'e pattern of graphene on bulk {\it h}-BN as a function of angle as calculated by Eq.~\ref{eq.length}
 The equilbrium lattice mismatch $\sigma$ is 1.8\%, whereas 1.7\% corresponds to a slight graphene stretching.}
 \label{fig.Lvstheta}
\end{figure}

Experimentally, for graphene ($a_C^{exp}$\,=\,1.4197) on bulk {\it h}-BN ($a_s^{exp}$\,=\,1.4460) it is $\rho$\,=\,1.018.
The  sound velocity ratio $c_L/c_T$\,$\approx$\,1.6 for graphene (see e.g.\ Ref.~\citenum{koukaras2015phonon}).
This means $\delta$\,=\,1.56, leading to a predicted theoretical misalignment by $\theta_{NM}$\,$\simeq$\,$\pm$0.45$^\circ$.
A lattice misalignment smaller than 1$^\circ$ may appear hard to detect, but the much larger
moir\'e pattern rotation angle $\psi$, satisfying\cite{yankowitz2012}

\begin{equation}\label{eq.moire}
 %\cos(\theta) = \rho^{-1}\sin^2(\psi) + \cos(\psi) \left[1-\rho^{-2} \sin^2(\psi)\right]^{1/2} ,
 \tan \psi = \frac{\sin \theta }{\rho-\cos \theta},
\end{equation}

will yield $\psi$\,$\gg$\,$\theta_{NM}$ much easier to visualize, effectively employable as a magnifying lensing.
Moreover, for a general $\theta$ the adsorbed graphene and the incommensurate substrate lattices form a moir\'e coincidence pattern
of length $L$,\cite{yankowitz2012,hod2016}

\begin{equation}\label{eq.length}
 L = \frac{a_C \sqrt{3}(1+\sigma)}{\sqrt{2(1+\sigma)(1-\cos \theta)+\sigma^2}}~~,
\end{equation}

where $\sigma$\,=\,$(a_s-a_C)/a_C$\,=\,$\rho-1$.  The Novaco-McTague misaligned state of graphene $\theta_{NM}$\,$\simeq$\,0.45$^{\circ}$
would imply a moir\'e pattern of length $L$\,=\,12.4\,nm and moir\'e rotation angle $\psi$\,=\,$\pm$22.9$^\circ$.
Simulated moir\'e patterns of graphene on {\it h}-BN (see Method) at $\theta$\,=\,0$^\circ$, 0.45$^\circ$
and 1.5$^\circ$, are compared in Fig.~\ref{fig.moire}.
The decrease of $L$ for graphene/{\it h}-BN obtained through Eq.~\ref{eq.length} is shown as a function
of $\theta$ in Figure \ref{fig.Lvstheta}.

Experimentally, moir\'e patterns with a length of approximately 14\,nm have been reported,\cite{zhang2015, woods2014commensurate, tang2013precisely}
consistent with perfect alignment, $\theta$\,=\,0$^\circ$, and in disagreement with $L$\,=\,12.4\,nm predicted by
the above theory. It was also observed that after long annealing,
micron sized graphene flakes initially at different angles slowly rotated towards $\theta$\,$\approx$\,0$^\circ$,
the slow kinetics indicating a flat dependence of energy on angle for $\theta$\,$<$\,0.7$^\circ$.\cite{woods2016macroscopic}
In general, it is likely that the observed deposition patterns could be out of equilibrium. The moir\'e lengths might also depend on graphene stretching,
which if present would decrease the lattice mismatch $\sigma$. Figure \ref{fig.Lvstheta} shows that the length of the moir\'e pattern
at 1.8\% mismatch and $\theta$\,=\,0$^\circ$ is close to that of the marginally smaller 1.7\% mismatch (globally stretched graphene layer) and
$\theta_{NM}$\,$\simeq$\,0.45$^\circ$. Our present goal is to establish a more precise theoretical understanding of equilibrium alignment to be
expected for unstretched graphene on {\it h}-BN.

\section{Method}

For our simulations we modeled the graphene/{\it h}-BN system as a fully mobile single layer graphene on a flat {\it h}-BN
monolayer substrate whose out-of-plane motion was inhibited, while in-plane motion was allowed.
Even if in reality {\it h}-BN is not vertically rigid, its top layer, resting on
the semi-infinite lattice underneath (that would be much more cumbersome to simulate),
has a substantially smaller flexibility than the graphene membrane.
The interatomic interactions within the graphene and {\it h}-BN layers were described by an optimized Tersoff potential ($a_C$\,=\,1.439\,\AA, $a_s$\,=\,1.442\,\AA).
\cite{lindsay2010optimized,cagin2012}
Graphene interactions with {\it h}-BN  were described with
the Kolmogorov Crespi (KC) potential\cite{KC2005} modified as described in Ref.~\citenum{slotman2015effect}, where the strength of the KC potential
was doubled for C--N interactions and reduced to 60\% for C--B ones.
Also, since the C--C bond length depends on the chosen graphene potential, we adopted a slightly rescaled planar simulation cell size
so that the graphene/{\it h}-BN size ratio exactly matches the experimental ratio $\rho$.
A sequence of 21 unit cells was created describing graphene adsorbed on {\it h}-BN with increasing misalignments angles from 0 to 30 degrees,
especially focusing on the small angles. For each angle we constructed a sample as in Ref.~\citenum{slotman2015effect} and carefully minimized its
classical energy (T\,=\,0), by allowing all atoms to relax their positions, while keeping at the same time the chosen overall alignment
angle $\theta$ blocked by the periodic boundary conditions.

A convergency test on the graphene corrugation $\Delta$z as a function of the number of the underlying {\it h}-BN layers, showed
unreasonably large vertical {\it h}-BN displacements when vertical mobility was allowed for even up a dozen layers.\cite{hod2016} Eventually,
in the semi-infinite the vertical displacements would heal out: but that limit is very far away. In line with that, the simple assumption
of a z-rigid, in-plane mobile {\it h}-BN substrate, adopted in the rest of this work, immediately provided very good agreement with
the experimentally determined geometry (see Fig.~\ref{fig.moire}: $\Delta$z\,$\simeq$\,40\,pm vs.\ $\Delta$z$^{exp}$\,$\simeq$\,35\,pm).\cite{zhang2015,LeRoy2011}
As it turned out, the two crucial elements that influenced alignment or misalignment were graphene corrugation and in-plane deformability of  {\it h}-BN.
To understand the role of out-of-plane motion of graphene, easily permitted by its soft ZA modes, we repeated all calculations by
imposing a solidal motion of graphene atoms in the out-of-plane direction, which blocked corrugations. To investigate the effect of {\it h}-BN in-plane
mobility, we also considered the case of a fully rigid {\it h}-BN plane. By the combination of the above cases we were able to extrapolate
the contribution to energetics of some of the fundamental degrees of freedom involved.

\begin{figure}[b!]
  \centering
  \includegraphics[width=0.7\columnwidth]{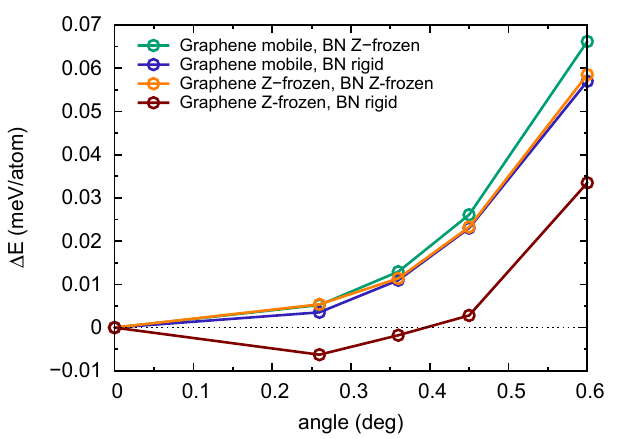}
  \caption{Variation of the total energy $E_{tot}$ as a function of the misalignment angle $\theta$ between graphene and {\it h}-BN substrate .
  Each curve is obtained by considering different constraints as specified in the legend (see Method).}
  \label{fig.Evstheta}
\end{figure}

\section{Results }

The results of Fig.~\ref{fig.Evstheta} show the resulting angle-dependent changes of the total energy $\Delta E$, obtained by considering
the interplay between the intra-graphene (elastic) energy $E_{intra}$, and the graphene/{\it h}-BN (adhesive) interlayer contribution $E_{inter}$.
The total energy profile is very flat up to 0.26$^\circ$ and, contrary to theoretical expectations, there is no well-defined minimum
at or near $\theta_{NM}$.

In conclusion, simulations show that misalignement does not really occur. While that is compatible with several observations, it contradicts Novaco-McTague
theory, which ought to have been applicable to this case. We must clarify why.

\begin{figure}[b!]
  \centering
  \includegraphics[width=0.7\columnwidth]{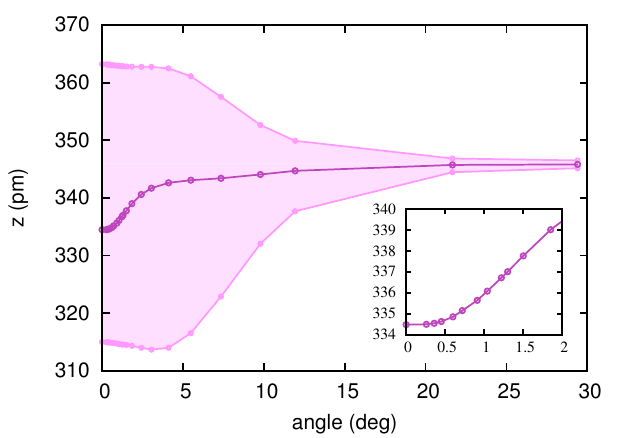}
  \caption{Minimum, maximum and average value of the $z$ coordinate of carbon atoms in graphene over the flat {\it h}-BN surface plane as a function of
  misalignment angle $\theta$. The wide spread is due to the large graphene corrugation. Inset: average values in the 0--2$^\circ$ range.}
  \label{fig.zvstheta}
\end{figure}

\begin{figure}[b!]
  \centering
  \includegraphics[width=0.49\columnwidth]{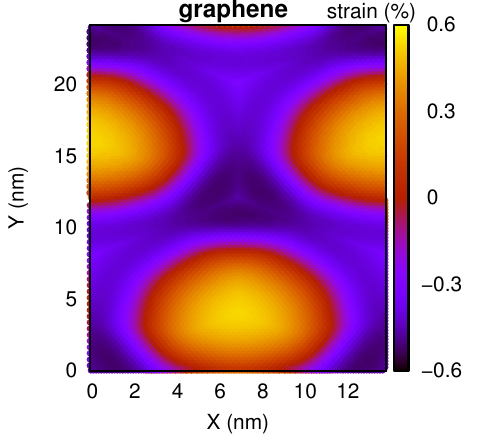}
  \includegraphics[width=0.49\columnwidth]{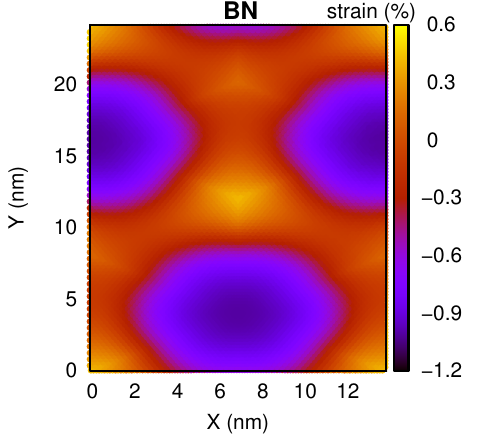}\\
  \includegraphics[width=0.49\columnwidth]{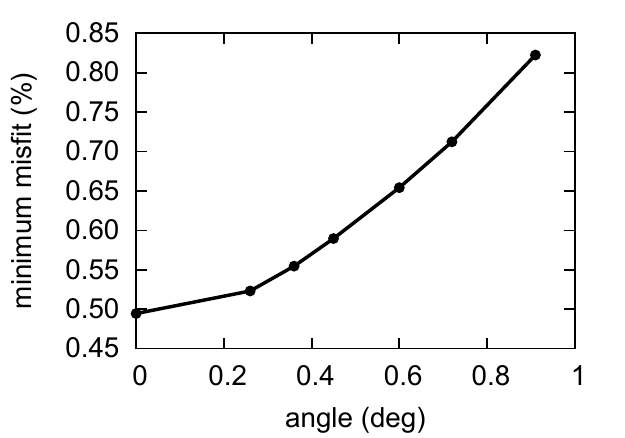}
  \caption{(top panels) Bond strain map for a mobile graphene deposited on a Z-frozen {\it h}-BN layer at $\theta$\,=\,0$^{\circ}$;
    (bottom panel) smallest calculated misfit between graphene and {\it h}-BN bond lengths as a function of the misalignment angle $\theta$;
    misfit value at each angle is obtained by ($l_{BN}^{min}$\,-\,$l_{CC}^{max}$)/$l_{BN}^{min}$, where $l_{BN}^{min}$ is the
    minimum bond length in the {\it h}-BN layer, while $l_{CC}^{max}$ is the maximum bond length in graphene. }
  \label{fig.bond_stress}
\end{figure}

The crucial clues are provided by structure: equilibrated graphene does not lie flat on {\it h}-BN. Fig.~\ref{fig.zvstheta} shows the range of z-distances between
individual carbons in graphene and the rigid {\it h}-BN substrate plane.
At the same time, and equally important, the BN planar lattice does not remain unperturbed, but to some extent mirrors the moir\'e.
The vertical corrugation displacements of graphene over the substrate are as large as
$\pm$8\% near $\theta$\,=\,0. In the Novaco theory, strictly 2D, the monolayer at $\theta$\,=\,0 has a higher energy than $\theta$\,=\,$\theta_{NM}$.
However, graphene as a flexible membrane is free to relax in the third, vertical direction. The vertical relaxation will reduce the energy, both for $\theta$\,=\,0
and $\theta$\,=\,$\theta_{NM}$, but the two energy gains need not be the same. Because at $\theta$\,=\,0 the misfit solitons are longitudinal and initially
cost more energy than the shear misfit solitons at  $\theta_{NM}$, it is natural that vertical relaxation will gain more energy than that at $\theta_{NM}$.
The result is that in general the Novaco-McTague rotation is weakened by vertical corrugation, and can therefore even disappear depending on actual numbers.
The vertical corrugation is accompanied by a nontrivial in-plane distortion of BN. As shown in Fig.~\ref{fig.bond_stress}, the BN lattice squeezes into
quasi-commensurability in the regions where graphene and BN adhere closely,\cite{woods2014commensurate} and can release back compensating its strain
in the soliton regions where graphene bulges outwards. The combined result of the vertical graphene corrugation and of the concurrent in-plane BN
lattice modulation is to eliminate the Novaco-McTague misalignment.

As a decisive step to verify this hypothesis we repeated the simulations by keeping graphene artificially flat,
allowing only in-plane relaxations and impeding corrugations, while also keeping the {\it h}-BN planar lattice fully rigid .
Once we fulfil in this manner all the ideal Novaco-McTague conditions, we indeed recover, as shown in Fig.~\ref{fig.Evstheta},
a small but nonzero equilibrium rotation of about 0.26$^\circ$. That confirms that vertical corrugations of the graphene monolayer,
along with a matching in-plane strain pattern of the {\it h}-BN substrate are responsible for the weakening and essential
suppression of misalignment on {\it h}-BN, that would otherwise be expected from the flat Novaco-McTague theory.

There is to our knowledge no direct modification of the Novaco-McTague formulation that would theoretically describe the reason
why corrugation on the one hand and the accompanying modulation of in-plane substrate strain on the other hand reduce and
eventually eliminate the tendency to misalign the graphene lattice over the substrate. However, the simple structural analysis
makes the physical reasons clear enough.
With direct reference to the moir\'e shown Fig.~\ref{fig.moire}, the graphene/{\it h}-BN epitaxy comprises two regions: the closely
adhesive hexagons, and the vertically corrugated soliton lines where graphene detaches from the substrate. As mentioned, the detachment
reduces -- screens, as it were -- the cost of the soliton. That reduction will be larger for the very costly longitudinal solitons in the
$\theta$\,=\,0 case than for the cheaper shear solitons at $\theta$\,$>$\,0 favoring alignment. On the other hand, the in-plane {\it h}-BN
strain brings the two lattices locally closer to commensurability, with a lattice mismatch reduction from 1.8\%, down to nearly zero,
in agreement with experiment.\cite{ woods2014commensurate} The  larger size moir\'e  at $\theta$\,=\,0  implies larger locally commensurate
hexagons, inside which the graphene/{\it h}-BN adhesion is stronger.  As a result, both factors conspire to stabilize the aligned state
$\theta$\,=\,0.

\section{Friction}

The next and last point of our concern is the sliding friction of graphene on incommensurate {\it h}-BN.
Friction is an important property with respect to anchoring and moving one system with respect to the other.  In our case, we examine the question how much would sliding friction be influenced by either alignment or small misalignments, whichever the case.
We do not try to address static friction, which is important but hard to study experimentally, and also hard to pursue computationally given the very large supercells required to reduce the finite-size effects, which are especially relevant at low velocities.
Actually, in a structurally lubric (``superlubric'') system like the present one, the dynamic friction per unit area at infinite size is expected to be proportional to the velocity $v$, therefore always larger than static friction, the latter arbitrarily small in the $v$\,=\,0 limit.
To obtain dynamic friction, we simulated the sliding of graphene through
non-equilibrium molecular dynamics, by applying a force $F$\,=\,0.001\,meV/\AA\ to each C atom in the $x$-direction (perpendicular to the B--N substrate bond), typically producing a center-of-mass speed of 0.5\,m/s.
Application of a sliding force to the-center-of-mass is not too different from pushing graphene by means of a large on-top AFM tip, while it does differ form the action of a tip exerting a side pushing on a bordered system.
Although this speed is large, the viscous-like proportionality between speed and friction typical of a lubric situation like ours suggests that relative results would not change at lower speed regimes, harder to access in simulation.

The frictional heat was absorbed through a Langevin damping $mv\gamma$ where $\gamma$\,=\,0.1\,ps$^{-1}$ and $m$,$v$ are the C-atom mass
and velocity, respectively. We conducted frictional simulations at $\theta$\,=\,0, %0.26,
 0.45, and 1.5$^\circ$, for a fully mobile graphene over a rigid or a Z-frozen {\it h}-BN substrate.
The average dissipated frictional power (per C atom) was evaluated by\cite{vanossi2012}

\begin{equation}\label{friction}
 p_{fric} = \overrightarrow{F} \cdot <\overrightarrow{v_{cm}}> - \gamma m < |\overrightarrow{v_{cm}}|^2 >
\end{equation}

where $v_{cm}$ is the center-of-mass velocity along x of the N atoms in the graphene sheet.
The frictional power obtained are reported in Table~\ref{tab.friction}.
For the more realistic case of mobile {\it h}-BN substrate the sliding friction generally remains constant up to about 0.45$^\circ$, where the energy landscape and the interface strain map remain practically unchanged due to C-BN squeezing into quasi-commensurability (see Fig.~\ref{fig.bond_stress}).
Conversely, moving towards larger misalignment angles, e.g.\ 1.5$^\circ$, dissipation experiences a singnificant drop resembling the tribological response of almost incommensurate interface, as proved by the more sinusoidal profile of Fig.~\ref{fig.moire}, right panel.
Due to the toughness of squeezing into such quasi-commensurability for graphene over a rigid substrate, we clearly see that the frictional values obtained in this case are much less influenced by the misalignement angle. The increase in the measured friction due to substrate mobility is much more pronounced at 0-0.45$^\circ$ ($\sim$+50\,\%) than for the misaligned system at 1.5$^\circ$ ($\sim$+25\,\%).

\setlength{\tabcolsep}{6pt}
\begin{table}
\begin{center}
  \begin{tabular}{|l|cc|}
    \hline
    $\theta$   &  BN rigid     & BN Z-frozen \\
    \hline
%     0.0        & 1.270 (0.996) & 1.999 \\
%    % 0.26       & 1.443 (1.439) & 1.793 \\
%     0.45       & 1.413 (1.412) & 2.002 \\
%     1.5        & 1.228 (1.210) & 1.532 \\
    0.0        & 1.270  & 1.999 \\
    0.45       & 1.413  & 2.002 \\
    1.5        & 1.228  & 1.532 \\
    \hline
  \end{tabular}
\end{center}
\caption{  $p_{fric}$ (in units of 10$^{-6}$ meV/ps) at $\theta$\,=\,0, %0.26,
 0.45, and 1.5$^\circ$, obtained by Eq.~\ref{friction} (see text) for a fully mobile graphene over a rigid or a Z-frozen {\it h}-BN substrate.
% The averages in Eq.~\ref{friction} can be taken over time, $\frac{\int v(t)dt}{\int dt}$, or in space, $\frac{\int v(x)dx}{\int dx}$. When the two averages differ, the $p_{fric}$ obtained with spatial average is reported in parenthesis.
}\label{tab.friction}
\end{table}

%%%%%%%%%%%%%%%%%%% CONCLUSIONS %%%%%%%%%%%%%%%%%%%%
\section{Conclusions}

In summary, we found that the spontaneous corrugation due to vertical $z$-relaxations of the adsorbed graphene monolayer
with an associated in-plane strain pattern of the {\it h}-BN substrate leading to locally quasi-commensurate portions of the
incommensurate  moir\'e are the elements that remove the small misalignment  predicted by classic flat-adsorbate over
rigid-substrate epitaxial theory.
The closeness in energy between the aligned and the slightly misaligned geometries is in turn reflected by not too dissimilar sliding
frictions obtained for aligned or misaligned lattices. On the other hand, overlayer corrugation and substrate strain
bring about an increase of sliding friction, the better interdigitation of the two lattices reflecting in a better anchoring between the two.
The above information is of  importance for a correct enginering of the graphene/{\it h}-BN bulk alloy, which thanks to its
structurally lubric interface could combine flexibility with extreme strength, suitable for flexible coating applications,\cite{lee2014science}
high performance cables,\cite{landi2011nanoscale,tai2016apl} and probably more. The type  of interplay between corrugation and strain
which we found to be responsible for stabilization of the aligned state will also bear consequences for electronics
applications, as it is likely to influence transport in a nontrivial manner. More generally, it can be expected that this
physical picture, once modified to account for different parameters, will be relevant to all strong layer sheet deposited on crystalline substrates.

\section*{Acknowledgements}

{\footnotesize We acknowledge COST Action MP1303. Work in Nijmegen is part of the research program of the Foundation for Fundamental Research on Matter (FOM),
Netherlands Organisation for Scientific Research (NWO), and funding from the European Union Seventh
Framework Programme under grant agreement No.\ 604391 Graphene Flagship. Work in Trieste was carried out under the ERC Advanced Grant No.\ 320796-MODPHYSFRICT. Discussions with D.\ Mandelli, M.\ Urbakh, and O.\ Hod are gratefully acknowledged.}

\end{document}